**From institutional journalism to political extremes: Structural asymmetries in YouTube's news recommender during the 2022 French presidential campaign**

**Julien Figeac**

Research Director

UTOPI

Centre National de la Recherche Scientifique (CNRS)

0000-0001-8324-8842

**Erwan Le Merrer**

Researcher

ARTISHAU

Institut national de recherche en sciences et technologies du numérique (INRIA)

0000-0001-8344-2135

**Marie Neihouser**

Associate Professor

CESSP

Université Paris 1 Panthéon-Sorbonne

0000-0003-0685-9599

**Gilles Tredan**

Research Director

LAAS

Centre National de la Recherche Scientifique (CNRS)

0000-0003-4473-4332

**Corresponding author:** Julien Figeac, Julien.figeac@cnrs.fr , CNRS, UTOPI, Toulouse, France.

**From institutional journalism to political extremes: Structural asymmetries in YouTube's news recommender during the 2022 French presidential campaign**

## Abstract.

Debate has intensified over how algorithmic recommendations shape news visibility, particularly regarding structural and ideological bias. This study examines YouTube's News Recommender System (NRS) in the French "News" section, using a large-scale audit based on automated browsing agents operating under cold-start conditions, without prior viewing history, to isolate the algorithm's structural baseline independently of user preferences.

Our results reveal that YouTube's NRS operates within a structurally asymmetric visibility regime: even in the absence of prior user behaviour, recommendation outputs disproportionately privilege platform-native creators and politically engaged content over established journalism. Television channels retain some relative advantage, but institutional news outlets, particularly press and radio, are systematically marginalised across recommendation sequences. Among political content, figures affiliated with far-left and far-right movements occupy a disproportionately visible position in recommendation outputs. Rather than correcting this asymmetry, continuity effects reproduce and stabilise it across recommendation sequences. These results show that visibility in YouTube's News section cannot be reduced to accumulated user preferences alone, but is also shaped by a structural bias embedded in the platform's recommendation system, which plays an important role in sustaining the prominence of political actors affiliated with the far left and far right.

Understanding how news recommender systems (NRS) shape information visibility has become a central challenge for both communication research and platform governance. In the European context, the *Digital Services Act* requires very large platforms to assess the systemic risks associated with their recommendation systems, including their potential effects on civic discourse and media pluralism. This regulatory framework reflects a dual analytical demand: on the one hand, understanding whether recommendation sequences amplify or reduce specific political and informational configurations in content exposure; on the other, documenting the baseline distribution of visibility that recommendation systems produce independently of accumulated user preferences.

Most existing audit research addresses the first of these questions. Large-scale studies have examined how YouTube's recommendation system shapes ideological exposure through sequential recommendations, revealing congruence effects and differential amplification depending on users' prior consumption profiles (Haroon et al., 2023; Brown et al., 2022). Recent work has even concluded that algorithmic recommendations play a moderating rather than radicalising role, with individual preferences explaining most of the variance in partisan consumption (Hosseinmardi et al., 2024; Yu et al., 2024). Yet these studies share a common design feature: they analyse recommendation dynamics as experienced by users with established viewing histories. As a result, they can assess how the algorithm interacts with particular user histories or counterfactual exposure scenarios, but they provide less leverage for identifying how visibility is distributed prior to any personalisation across different user trajectories.

YouTube provides a particularly relevant case for examining both dimensions of this problem. As a platform combining mainstream media, independent creators, political figures, and alternative news sources, it structures content visibility through algorithmic ranking rather than editorial curation (Evans et al., 2023; Bodó, 2021). Unlike most social media platforms, YouTube features a dedicated "News" section curated by an algorithmic system, making it possible to observe how recommendation infrastructures organise news visibility across these different types of actors. This is especially important for the political dimension of our analysis, since the prominence of political figures in recommendation outputs can be examined more precisely in the French case, where the multi-party structure of the political field, spanning from far left to far right, provides a more differentiated setting than the bipartisan contexts in which most existing YouTube audit studies have been conducted. It therefore allows a finer-grained assessment of the direction and structure of ideological asymmetries in recommendation outputs.

In this context, this study investigates how YouTube's NRS structures news visibility within the platform's "News" section in the French case. Using automated agents operating under cold-start conditions, without prior viewing history, we first document the baseline distribution of visibility at step 1 of the recommendation sequence, providing an empirical approximation of how the algorithm hierarchises content in the absence of accumulated user preferences. Unlike counterfactual approaches that estimate what recommendation systems change relative to an alternative exposure scenario, our cold-start design documents the initial distribution of visibility presented by the system before accumulated user preferences intervene. We then analyse how recommendation sequences from steps 2 to 5 interact with this initial distribution, assessing whether continuity effects reproduce, amplify, or correct it. Our analysis examines two key dimensions: the types of media sources promoted (institutional outlets vs. YouTube-native creators) and the ideological orientation of content produced by political figures appearing in recommendation outputs.

## Theoretical background

### *The platformisation of news on social media: reconfiguring media hierarchies*

Unlike traditional media institutions, which operated through editorial gatekeeping, social media increasingly rank content according to engagement metrics (Bodó, 2021). This shift, referred to as platformisation (Nieborg & Poell, 2018), has introduced new criteria for distributing visibility, structured around platform logics and commercial imperatives rather than journalistic norms (Evans et al., 2023). In this context, YouTube constitutes a particularly important case. Unlike platforms such as TikTok, where proactive algorithmic news distribution appears limited, it features a dedicated algorithmically curated “News” section (Hagar & Diakopoulos, 2025), in which institutional media actors compete alongside YouTube-native creators, political figures, and other non-traditional sources within the same recommendation space (Bodó, 2021; Kalogeropoulos et al., 2018). Research has shown that algorithmic ranking can privilege some sources, formats, and political content (Markmann & Grimme, 2021). Consequently, visibility depends less on editorial authority than on compatibility with platform ranking logics, creating conditions under which platform-native creators and influential politicians may gain visibility at the expense of established news organisations (Hase et al., 2023).

Within this environment, not all actors are equally positioned to benefit from these ranking logics. Traditional news organisations, particularly in print and radio, often struggle to adapt to the platform's interactive features, while independent creators more readily exploit engagement-driven formats (Nieborg & Poell, 2018). More broadly, research has shown that YouTube's algorithm favours content that maximises user engagement, including politically divisive or controversial material (Faddoul et al., 2020; van der Breggen et al., 2025). YouTube's autoplay function has been described as potentially guiding users towards more extreme content through incremental exposure pathways (Brown et al., 2022), and large-scale audits have shown that extremist content can become reachable through recommendation pathways, including in settings that do not account for personalisation (Ribeiro et al., 2020). At the same time, recent work has cautioned against treating such radicalisation effects as systematic or uniform across user trajectories (Hosseinmardi et al., 2024). Taken together, these studies suggest recurring asymmetries in what becomes visible, yet most examine them as experienced by users with established viewing histories, leaving open the question of the system's baseline outputs prior to any personalisation.

While recommender systems can, in some cases, produce diverse outputs comparable to those of human editors (Möller et al., 2018), such assessments typically focus on the properties of specific algorithmic designs rather than the structural baseline from which recommendations emerge prior to any user interaction. These dynamics point to what we term a structurally asymmetric visibility regime (Gillespie, 2014; Sandvig et al., 2014): a configuration in which the recommendation system's default ranking logic produces an asymmetric distribution of visibility that precedes user interaction and is reproduced across recommendation sequences, systematically privileging engagement-driven, politically salient, and platform-native content over institutional journalism. While existing work has shown that user interaction patterns significantly shape recommendation outcomes (Hosseinmardi et al., 2024), it also makes platform-level structural effects harder to isolate from individually driven ones. Our cold-start design addresses this issue directly and allows us to document this structural asymmetry at the level of the algorithm's baseline operation, independently of individual user preferences. Our sequential analysis then examines whether continuity effects reproduce, amplify or correct this asymmetry across recommendation trajectories.

### *Platform logics and source selection in YouTube's News algorithm*

Within this structurally asymmetric visibility regime, source selection in YouTube’s News algorithm is shaped less by editorial criteria than by platform-specific ranking mechanisms tied to engagement signals. Recent audit studies have identified two main forms of algorithmic bias on YouTube: interest bias, which favours entertainment over public affairs (Yu et al., 2024), and ideological bias, which directs users towards like-minded and, at times, extreme political content (Huang & Yang, 2024). However, most of this work analyses the platform as a whole, without isolating curated environments such as the “News” section, where visibility depends on specific interaction signals, including watch time, likes, shares, and autoplay, rather than on editorial criteria alone.

In this environment, actors are not equally positioned to benefit from platform ranking logics. Established news institutions often struggle to adapt fully to these dynamics, as their production standards and economic models rely less on in-platform engagement than those of platform-native creators (Poell et al., 2019). On YouTube, they frequently retain one-way broadcasting routines and invest less in interactive formats that generate visibility within the platform, such as Shorts, live discussions, or audience participation features. By contrast, independent YouTube-native creators, including citizen journalists and political figures, tend to produce content that is more directly aligned with YouTube's ranking environment, both in terms of format and interactive potential (Nieborg & Poell, 2018). Studies of YouTube's recommendation system further indicate that videos associated with high user engagement metrics are more likely to be promoted, irrespective of their journalistic credibility (Faddoul et al., 2020). This aligns with previous findings on digital journalism, where interactive formats drive greater algorithmic amplification (Bodó, 2021). Given YouTube's emphasis on interaction-based ranking, we therefore hypothesise that content creators who natively optimise their videos for these platform-specific dynamics are favoured over those who do not.

*H1a. YouTube's algorithm promotes, within the News Section, content produced in formats more compatible with the platform's ranking environment, particularly content from YouTube-native creators.*

There is a broader tendency to favour content that is better aligned with platform-native formats and the platform's engagement-driven logics, and this is not limited to YouTube-native creators. Similarly, Hase et al. (2023) argue that media outlets producing visual content may have a structural advantage as social media platforms increasingly centre on or algorithmically favour audiovisual content. Building on Jenkins and Deuze (2008) concept of convergence culture, one could hypothesise that audiovisual formats originally designed for television, since they share key structural features with user-generated content on YouTube, are more likely to be favoured by the platform's NRS. In contrast, traditional print and radio journalism, whose formats are less compatible with video platforms, face greater challenges in adapting to YouTube's features. This difference highlights how legacy media sectors are unevenly positioned to engage with platform logics.

*H1b. YouTube's algorithm promotes content produced by television channels more than print- or radio-based journalism, insofar as its audiovisual format is more compatible with the platform's ranking environment.*

However, platform logics do not operate independently of user behaviour. YouTube's NRS adapts its recommendations to viewing patterns and interaction habits (Karimi et al., 2018; Feng et al., 2020; Hosseinmardi et al., 2021; Yu et al., 2024), creating a feedback loop in which algorithmic outputs shape consumption, and consumption in turn conditions future outputs (Le Merrer et al., 2023).

In this context, the weight given to platform-native formats in recommendations may vary depending on users' previous engagements. Users who regularly consume audiovisual news content, such as television news segments, are more likely to receive similarly formatted recommendations than video adaptations of print-based journalism. This suggests that user behaviour can modulate how content is ranked within the platform's engagement-based logic, even if the system's overall orientation remains platform-driven. While existing studies have primarily documented how accumulated cross-session viewing histories shape recommendation outcomes, our cold-start design allows us to isolate a more constrained dimension of this dynamic: within-session sequential effects, in which each viewed video constitutes the only behavioural signal available to the algorithm.

*H1c. YouTube's News recommendations are shaped by prior exposure to news content, with previously viewed videos guiding subsequent recommendations and channelling users along specific recommendation trajectories.*

### *Political visibility and ideological trajectories in YouTube's recommendation system*

Beyond source hierarchies, platform logics also shape the visibility of political actors. If engagement-driven ranking favours YouTube-native formats, it may also redistribute visibility unevenly across political orientations. The question becomes not only whether political content is amplified, but which ideological trajectories are reinforced through sequential recommendation dynamics.

Research suggests that algorithmic recommendation systems do not treat all forms of opinion-based content equally, but rather tend to favour content that is emotionally charged, divisive, and ideologically polarising. On social media more generally, content that reinforces partisan identities or sparks controversy is more likely to elicit strong user reactions and sustained attention, thereby increasing its visibility within algorithmic recommendation systems (Brown et al., 2022). Some studies conducted in the French context, specifically on YouTube, suggest that algorithmic systems may guide users gradually towards more radical or emotionally provocative content, a phenomenon often described as "rabbit hole" dynamics (Faddoul et al., 2020; van der Breggen et al., 2025). While these studies document the amplification of divisive content on YouTube, they do not examine the visibility of specific political actors.

Most of the audit studies specifically examining ideological dynamics on YouTube have, however, been conducted in the United States, where the bipartisan structure of the political field constrains ideological classification to a binary left-right axis (Haroon et al., 2023; Ribeiro et al., 2020). By contrast, research conducted in the French context on Facebook and Twitter has shown that far-right actors play a central role in the circulation of unreliable content online (Figeac et al., 2020; Figeac et al., 2021), and that extreme right-wing groups have long utilised digital platforms strategically to enhance their visibility (Caiani & Parenti, 2016). The multi-party structure of the French political system, with its five distinct ideological positions from far left to far right, thus provides a more analytically differentiated setting, making it possible to assess whether YouTube's algorithm favours a specific political orientation or, more broadly, content associated with ideological distance from the political mainstream, a distinction that binary left-right operationalisations structurally foreclose.

*H2a. YouTube's algorithm is more likely to recommend content produced by political figures affiliated with the parties at the ideological extremes, as their communication strategies often align with the platform's engagement-based ranking logic.*

Rather than being solely guided by the institutional status of content producers, YouTube's NRS seems to organise recommendations based on patterns of engagement-driven amplification. Previous studies have demonstrated that recommendation algorithms are highly sensitive to user behaviour, tailoring suggestions in response to individual viewing histories (Karimi et al., 2018; Feng et al., 2020; Hosseinmardi et al., 2021), reinforcing topical preferences over time. Recent work also shows that recommendation systems can exert a stronger influence on users' exposure than users' exposure has on subsequent recommendations, and that balanced news input can reduce ideological biases in YouTube recommendations (Yu et al., 2024). These reinforcement mechanisms may contribute to increasingly narrow content trajectories (Faddoul et al., 2020).

*H2b. Within the News section, the ideological orientation of political content recommended to users is shaped by prior exposure to political content, with previously viewed videos guiding subsequent recommendations and channelling users along politically aligned content trajectories.*

## Methodological approach

This study investigates how YouTube's NRS structures news visibility in the French "National News" section, based on a three-part methodological protocol: (1) the simulation of user behaviour via automated browsing agents operating under cold-start conditions; (2) the manual labelling of recommended content by source type and political orientation; and (3) the statistical modelling of recommendation dynamics across exposure sequences.

### *Simulating user behaviour through automated bots*

To investigate the dynamics of YouTube's NRS during the 2022 French presidential campaign, we conducted an automated audit using scripted browsing agents, also known as sock puppets, designed to simulate user behaviour in a controlled way. The audit ran continuously from January 17 to April 10, 2022, covering the final stretch of the campaign up to the first-round election day, across three parallel machines executing 180 sessions per day. Although not an experiment in the classical sense involving randomized treatments or human subjects, this audit protocol allows for the systematic observation of algorithmic outputs. Figure 1 illustrates the entire sequence followed by the bots, from session initiation and random video selection ($w_1$), to autoplay-driven navigation ($w_2$–$w_5$), and parallel recommendation capture ($r_1$–$r_{20}$ at each step). In total, this procedure was repeated over 12 000 independent browsing sessions, each reproducing the five-step sequence described in the figure, yielding approximately 59 000 watched videos and 1.31 million recommended ones.

**Figure 1. Audit protocol with sock puppet agents on YouTube's News section**

The bots were implemented using the Python Puppeteer library to control Chrome browsers. Each browsing session was executed within a dedicated Docker container, instantiated and destroyed after completion to ensure full isolation. This setup eliminated persistent tracking mechanisms (e.g., cookies, local storage, fingerprinting) and allowed for clean measurement conditions. Each session began with a cold-start profile (Karimi et al., 2018), devoid of any browsing history or personalisation signals, in order to observe how content is prioritised from a neutral user state.

Each bot followed a five-step viewing path. It began by accessing the French "National News" section of YouTube's homepage in a cookie-free browser instance. A video was selected at random among those displayed and watched for a variable duration centered on five minutes (±10%). The bot then automatically followed four subsequent videos using YouTube's autoplay feature, which simulates passive consumption shaped by the platform's real-time recommendation engine. During autoplay steps, the bot did not interact with the interface in any way (no clicks, likes, scrolls, or search queries were simulated) so as to isolate the influence of the default recommender path. During steps 2 to 4, bots extracted about 20 right-column recommendations per step. These videos were not watched, but their metadata were collected to capture what the algorithm proposed alongside each exposure.

### *Manual labelling of recommended sources*

To analyse the nature of the content recommended by YouTube's NRS, we performed a manual labelling of the collected sources, focusing on domain names and video metadata. We manually coded two dimensions:

1. *Source typology*. We categorised the top 1,000 domain names by media type: television channels, radio stations, print press, online-native news outlets, and YouTube-native content creators. An "Unknown" category was assigned to sources that were not included in the manual classification and could therefore not be assigned to any category within the established typology.

2. *Political orientation of content creators.* Within the category of YouTube-native content creators identified in the source typology, we assigned a political orientation (far left, left, centre, right, far right, neutral) only to those political figures whose public affiliation to identifiable political parties could be established. This coding follows prior classifications of the five main French political groups in research on online political communities (Figeac et al., 2020; Figeac et al., 2021). Only elected officials or clearly affiliated personalities were assigned an ideological label; figures affiliated with centrist parties were classified as “centre”, while unaffiliated or ambiguous cases were classified as “Neutral”.

Given the clarity and public visibility of the sources and content types, no inter-annotator agreement index was computed. The classification relied on information that is both factual and publicly verifiable: for instance, Le Monde was categorised as a print press outlet, while TF1 was classified as a television channel. Political orientation was determined based on the official affiliation of elected figures to well-established French parties, allowing for a clear positioning along the left–right ideological spectrum, including far-left and far-right.

These two classifications were applied to the top 1,000 most frequently recommended sources. Although these represent only 1.8% of the 55,697 unique domains in the dataset, they account for 89.6% of all recommendation occurrences (1.31 million), thereby covering the overwhelming majority of algorithmic exposure.

### *Statistical Modelling and Analytical Strategy*

To quantify how prior exposure shapes subsequent recommendations within YouTube’s News Recommender System (NRS), we conducted a series of logistic regression analyses. The unit of analysis is the individual recommended video. For each model, the dependent variable is a categorical indicator corresponding to one of the following: (1) the type of media source that produced the recommended video (TV, radio, print press, online-native, or YouTube-native), or (2) the ideological orientation of the political figure featured in the recommended video (far left, left, centre, right, far right, neutral).

Each model estimates the probability that a recommended video belongs to a given category, conditional on the attributes of the last watched video (namely its source type or political orientation). This allows us to assess whether YouTube’s NRS reinforces content similarity across exposure sequences, or instead promotes diversity in terms of media origin or ideological alignment.

All models were implemented as logistic regression models, with baseline intercepts included to account for overall recommendation frequency by category, so that the estimated coefficients capture the effect of prior exposure relative to this baseline. Results are visualised as log-odds heatmaps, illustrating the influence of prior exposure on subsequent recommendations across the two classification dimensions retained: source type and ideological orientation of political content.

This modelling strategy operates at two analytical levels. At step 1, recommendations are generated under cold-start conditions, with no prior viewing history available to the algorithm. The distribution of recommendations at this step therefore approximates the baseline visibility regime produced by the system independently of accumulated user preferences. From steps 2 to 5, each recommendation is conditioned on the type or political orientation of the previously watched video, allowing us to estimate whether continuity effects reproduce, amplify, or correct the asymmetries observed at the baseline. The combination of these two analytical levels makes it possible to distinguish the algorithm's default ranking outputs from the dynamics through which subsequent recommendations interact with this baseline.

## Results

Our results document YouTube’s NRS at two analytically distinct levels. The first section establishes the structural asymmetry of news visibility: drawing on cold-start data generated in the absence of prior user history, we examine how the algorithm distributes visibility across source types, showing a progressive marginalisation of institutional journalism and a corresponding rise of platform-native content. The second section examines the political dimension of this asymmetric regime by analysing how prior exposure shapes ideological continuity across recommendation sequences, and whether these dynamics contribute to the prominence of politically extreme actors within the recommendation landscape.

### *Eroding the legacy: Platform-native creators and the decline of institutional news*

We first examine the baseline distribution of source types at step 1, generated under cold-start conditions, before assessing how this initial configuration evolves across the subsequent autoplay steps (steps 2 to 5). Figure 2 traces how the distribution of recommended source types evolves from this cold-start baseline across the five stages of YouTube's recommendation process. It reveals how the platform's NRS and autoplay function interact to influence the visibility of various news categories within YouTube's "News" section.

**Figure 2. Evolution of source types across the five steps of the recommendation sequence**

At step 1, which corresponds to the cold-start baseline, television channels dominate with approximately 96,000 recommended videos. However, their visibility steadily declines throughout the trajectory, dropping to around 74,000 videos by the fifth stage. A similar trend is observable among other institutional outlets: print press recommendations fall from 24,000 to 8,000, online news websites from 27,000 to 18,000, and radio content from 63,000 to 50,000. These results show that traditional providers lose visibility at each step of the recommendation process, with a cumulative decline of −28.6% between the first and fifth steps.

By contrast, YouTube-native content (manually labelled videos produced by political figures and platform-native creators) follows the opposite trajectory. Starting at 51,000 recommendations, it increases to 65,000 by the fifth step, marking a relative growth of +27.5%. This pattern indicates that the algorithm increasingly favours content designed for YouTube's native ecosystem and optimised for its engagement metrics. Taken together, these results indicate that recommendation sequences from steps 2 to 5 do not correct the baseline asymmetry observed at step 1, but rather amplify it: institutional outlets become progressively less visible, while YouTube-native content gains in prominence.

This shift in recommendation patterns supports H1a, illustrating the progressive reorientation of visibility toward content adapted to YouTube's participatory and engagement-driven logic. The decline of traditional sources, in contrast to the growing prominence of YouTube-native content, reflects platformisation, where visibility depends on compatibility with the platform's engagement logic. These findings align with research highlighting the challenges institutional news organisations face in adapting to the algorithmic logic of digital platforms (Hase et al., 2023).

Lastly, the relatively greater persistence of television channels compared to other institutional outlets supports hypothesis H1b, which suggests that audiovisual formats are more aligned with YouTube's technical capabilities than print or audio-based journalism. Television content, due to its visual format and dynamic editing, is better suited to the platform's infrastructure and thus more likely to retain algorithmic visibility. Although television content also declines, its relative persistence compared to press and radio shows that audiovisual compatibility can partially mitigate algorithmic deprioritisation.

While Figure 2 outlines general trends in content distribution throughout the recommendation process, Figure 3 provides a more detailed perspective on how YouTube's NRS selects subsequent items based on the type of content most recently viewed. The analysis relies on a heatmap displaying the log-odds of a given source type being recommended following the viewing of another, allowing us to examine the influence of prior exposure on recommendation patterns (H1c).

**Figure 3. Log-Odds of recommended content types based on the last item watched**

*Note. Values represent log-odds from logistic regression models predicting the type of recommended item by the last watched source: Unknown ($R^2$ = 0.18; N = 130k), Press ($R^2$ = 0.08; N = 69k), Online ($R^2$ = 0.05; N = 109k), Radio ($R^2$ = 0.17; N = 277k), TV ($R^2$ = 0.10; N = 428k), YouTube ($R^2$ = 0.11; N = 303k).*

At first glance, the heatmap reveals a significant diagonal dominance, displaying high log-odds when the type of recommended source aligns with the type of the last watched video. This pattern is especially strong for radio (with a log-odds $\beta$ = 3.48) and press ($\beta$ = 2.98), indicating that the NRS tends to reproduce previous exposure when users engage with these types of content. Continuity is therefore strongest for these categories, supporting H1c.

However, a closer look at cross-recommendations reveals a more asymmetric dynamic, particularly in how YouTube-native content interacts with traditional news institutions. After watching a YouTube-native video, the next recommendation is most often of the same type ($\beta$ = 1.52), with traditional outlets consistently under-recommended (e.g., press −0.29, radio −0.59, TV −0.4). These patterns suggest that once users enter the YouTube-native stream, the algorithm rarely redirects them to traditional journalistic sources.

This observation helps to clarify the trend observed in Figure 2, where the visibility of traditional news institutions diminishes over time, while YouTube-native content steadily rises. The algorithm contributes to structurally orienting exposure toward sources natively aligned with YouTube's engagement logic. This mechanism supports H1a and H1b, illustrating how platform-native creators and audiovisual broadcasters benefit from structural compatibility with YouTube's recommendation logic. It also substantiates H1c, as the data show that once users engage with native YouTube content, the algorithm is unlikely to redirect them towards established news organisations, particularly those rooted in print or radio journalism.

***Polarised exposure: Visibility of far-left and far-right figures in recommendation outputs***

As with source types, we examine first the baseline distribution of political content at step 1, before assessing how this distribution evolves across steps 2 to 5. Figure 4 provides a detailed view of the visibility granted to political video creators according to their ideological affiliations. Many of these creators are not anonymous, but elected officials or public figures such as Jean-Luc Mélenchon (Far Left) or Florian Philippot (Far Right). Throughout the five stages of the recommendation process, content created by public figures affiliated with the far left dominates the recommendation landscape (444,116 recommended videos), with the far right content also occupying a prominent position (150,927 videos). In contrast, video creators associated with centrist (643 videos), moderate left-wing (480 videos), and moderate right-wing (292 videos) positions remain marginal throughout the recommendation process. This contrast highlights a key distinction: not all platform-native creators gain equivalent visibility. Political actors affiliated with far-left and far-right movements are disproportionately represented in recommendation outputs, a pattern consistent with the engagement-driven logics of the NRS. These results support hypothesis H2a. Crucially, this dominance is already established at the baseline (step 1): far-left and far-right content together account for the vast majority of politically labelled recommendations from the outset. While their visibility increases slightly across steps 2 to 5, centrist, left-wing, and right-wing content remains near-absent throughout the sequence, suggesting that the gap is structurally embedded rather than progressively constructed. The continuity effects measured in Figure 5 below confirm that the algorithm reproduces this asymmetric baseline distribution rather than offsetting it.

**Figure 4. Distribution of video recommendations by political leaning across the five steps of the recommendation sequence**

*Note: The vertical axis uses a logarithmic scale.*

In order to deepen this analysis, we assess the extent to which prior exposure to political content affects the ideological orientation of subsequent video recommendations (cf. Figure 5). These logistic regression analyses support H2b, as they reveal a pronounced homophilic pattern in YouTube's political recommendations: users are significantly more likely to be recommended content that aligns with the ideological orientation of the political videos they have previously watched. The highest coefficients are noted when users who have previously engaged with content associated with the left, the centre, or neutral positions are subsequently recommended videos from the same respective ideological category. This indicates a strong reinforcement for left-aligned content ($\beta = 7.19$), centrist content ($\beta = 8.56$), and neutral content ($\beta = 6.18$).

**Figure 5. Distribution of recommended political YouTube channels according to the political orientation of previously watched content**

Note. Values represent log-odds from logistic regression models predicting the political orientation of recommended YouTube channels as a function of the orientation of previously watched content: Unknown ($R^2 = 0.11$; N = 1.20 M), Far-Left ($R^2 = 0.15$; N = 88 k), Left ($R^2 = 0.40$; N = 436), Centre ($R^2 = 0.42$; N = 334), Right ($R^2 = 0.02$; N = 249), Far-Right ($R^2 = 0.11$; N = 15 k), Neutral ($R^2 = 0.53$; N = 8 k). Log-odds coefficients: *p < .05, **p < .01, ***p < .001. Exclamation marks (!) denote unusually low p-values, likely due to very small sample sizes in some ideological categories.

This finding warrants attention: in our manually labelled dataset of the 1,000 most recommended political channels, only 5 were classified as Left, 2 as Centre, and 2 as Neutral, compared to significantly larger counts for Far Left and Far Right creators. These high coefficients for underrepresented categories admit two interpretations: they may reflect stronger algorithmic reinforcement for rare ideological trajectories, or they may partly result from a constrained choice space, in which the very small number of channels within a category mechanically increases the probability of within-category recommendation. Given the small number of channels involved, these estimates should therefore be interpreted with caution. This interpretation becomes clearer when compared with the Far Left and Far Right categories, where the coefficients are comparatively lower ($\beta = 2.01$ and $\beta = 2.65$) despite much larger category sizes. This pattern suggests that reinforcement effects do not scale linearly with category prevalence in the recommendation data, but vary across ideological categories.

These findings refine H2b by showing that the algorithm does not merely reflect the ideological orientation of the previously watched content, but reproduces ideological coherence across successive steps and distributes visibility asymmetrically across ideological categories. For Far Left and Far Right content, which already dominates, continuity effects produce consistent but moderate reinforcement ($\beta = 2.01$ and 2.65), sustaining their structural prominence across successive steps. For underrepresented categories, the observed coefficients are substantially higher, though these estimates should be interpreted with caution given the small number of channels and the possibility of a constrained choice space effect. YouTube's NRS thus operates on two complementary levels: a structural asymmetry in the baseline distribution of visibility, reproduced and stabilised through continuity effects across recommendation sequences. While source-type asymmetries are progressively amplified across recommendation sequences, the prominence of far-left and far-right political content is driven primarily by the baseline distribution rather than by strong sequential reinforcement. Continuity effects sustain this initial asymmetry, but do not dramatically amplify it.

## Discussion

This study therefore documents a structurally asymmetric visibility regime in YouTube's "News" section. Our cold-start design shows that YouTube's NRS does not begin from a neutral baseline: even in the absence of prior viewing history, the algorithm privileges platform-native and politically engaged content over institutional journalism. This configuration can be understood as a form of structural bias embedded in the platform's recommendation system, one that precedes and shapes user behaviour rather than merely reflecting it.

This asymmetry becomes clearer when the two analytical levels of the design are considered together: step 1 reveals an asymmetric baseline distribution of visibility, while the sequential analysis from steps 2 to 5 shows that continuity effects reproduce and stabilise it rather than correcting it. In this sense, YouTube's NRS does not operate as a neutral conduit reflecting user preferences, but as a recommendation environment whose initial conditions already structure visibility asymmetrically (Gillespie, 2014). Our findings therefore complement prior work centred on user histories, interaction patterns, and personalised exposure (Karimi et al., 2018; Feng et al., 2020; Hosseinmardi et al., 2021) by showing that the trajectories users encounter are also shaped by these initial visibility conditions. At the same time, this asymmetry concerns the distribution of visibility within recommendation outputs. Our design, in line with audit-based approaches to algorithmic systems (Sandvig et al., 2014), allows us to identify this baseline asymmetry and to assess whether subsequent recommendation sequences reproduce, amplify, or correct it. What it does not determine by itself is whether this baseline distribution reflects, transforms, or amplifies the underlying supply of political and journalistic content available on the platform.

At the level of source type, the findings also show that YouTube's NRS privileges content structurally aligned with the platform's engagement logic, particularly audiovisual formats and engagement-oriented design. This favours YouTube-native creators whose production strategies match the platform's logic (H1a). In contrast, institutional outlets, especially print and radio outlets, appear disadvantaged, as their content aligns less closely with YouTube's optimisation dynamics. Television broadcasters, while experiencing a decline, retain some visibility due to stronger audiovisual compatibility (H1b). These findings show that content aligned with the platform's engagement logic favours YouTube creators (Nieborg & Poell, 2018) and marginalises established news organisations within algorithmic spaces (Hase et al., 2023).

At the level of political actors and ideological positioning, our findings support H2a by showing that politicians affiliated with far-left and far-right movements dominate the recommendation landscape by volume. Prominent individuals, such as Jean-Luc Mélenchon, maintain sustained visibility across recommendation sequences. Regarding H2b, the most robust finding is that continuity effects reinforce ideological alignment across recommendation sequences, including within Far Left and Far Right trajectories. The substantially higher coefficients observed for Left, Centre, and Neutral categories should be interpreted with caution, as they may partly reflect a constrained choice space linked to the very small number of channels available in these categories. Taken together, these results suggest that ideological recommendation on YouTube cannot be understood solely in terms of user-driven alignment, but must also be situated within a pre-structured distribution of visibility. These findings extend recent work showing that ideology-based recommendation contributes to political polarisation (Ludwig et al., 2025) by revealing that such asymmetry is already embedded in the platform's structural baseline, prior to any personalisation. Ideological alignment in the NRS is thus shaped not only by prior user behaviour, but also by an asymmetric baseline distribution of visibility that precedes any accumulation of individual preferences and is reinforced, rather than offset, by sequential recommendation dynamics.

These findings also complement existing audit studies conducted primarily in bipartisan contexts, where ideological asymmetries are typically assessed along a binary left-right axis. The multi-party structure of the French political field enables a more differentiated reading: rather than a directional bias favouring one political camp, the data reveal a bilateral pattern in which both far-left and far-right content gains disproportionate visibility, while centrist positions remain marginalised. This suggests that engagement-driven reinforcement may operate across ideological orientations rather than in favour of a specific one, a pattern that bipartisan operationalisations are less structurally equipped to detect.

Ultimately, these findings reflect wider concerns regarding the platformisation of journalism (Nieborg & Poell, 2018). By prioritising content aligned with its audiovisual and engagement-driven platform logic, YouTube's NRS reorders the distribution of visibility among news producers. In so doing, it redefines journalistic legitimacy, not in relation to editorial norms or institutional authority, but based on the extent to which content is compatible with the platform's engagement logic. This visibility, however, does not solely arise from platform dynamics; it also intersects with the broader media presence of certain actors, including those whose prominence predates the rise of digital platforms. More broadly, this suggests that the platformisation of news does not simply redistribute visibility among information sources, but reorders the criteria through which newsworthiness is recognised and sustained. This structural advantage in algorithmic visibility raises important interpretive challenges. While it often benefits ideologically polarising content, it is important to note that the algorithm does not assign political meaning per se. Rather, it promotes formats that maximise engagement, which often correlate with confrontational or partisan rhetoric. This dynamic is consistent with prior analyses of YouTube's recommendation system (Faddoul et al., 2020; Haroon et al., 2023).

More broadly, these findings clarify two distinct audit questions that platform governance should not conflate. Counterfactual and user-based approaches are designed to estimate whether recommendation systems amplify, moderate, or redirect users' trajectories relative to alternative exposure scenarios. Our cold-start design addresses a different, but equally important, question: it identifies the baseline distribution of visibility that a recommendation system imposes before accumulated user preferences intervene. This distinction matters directly for the regulatory logic of the European Union's Digital Services Act (DSA). If systemic risk assessment focuses only on personalisation effects, it may miss a prior and more structural layer of algorithmic bias: the default hierarchisation of actors and content at the point of entry. Our results illustrate this concretely: even prior to any personalisation, far-left and far-right political actors account for the overwhelming majority of politically labelled recommendations, while centrist and moderate actors remain structurally marginal. In the case of far-left and far-right content, continuity effects do not strongly amplify visibility. Rather, their disproportionate prominence is already embedded in the system's baseline outputs and is sustained, not corrected, across subsequent steps. When continuity effects then reproduce rather than correct this asymmetric baseline, the result is a recommendation environment in which the default visibility structure sustains the prominence of ideologically extreme actors across exposure sequences. From a DSA perspective, this constitutes a form of systemic risk that personalisation-focused audit frameworks alone are not designed to detect: not the amplification of individual trajectories, but the structural imbalance embedded in the system's default outputs. In this respect, the systemic risks associated with recommender systems concern not only what algorithms do to already-formed trajectories, but also the non-neutral visibility structures they establish from the outset.

### Limitations and Future Research

While this study offers new insights into the functioning of YouTube's NRS, several limitations must be acknowledged. Firstly, the analysis focuses on the French "News" section during the 2022 presidential election, a politically sensitive moment. Whether these patterns persist in less polarised contexts or other national settings remains an open question. The effects of algorithmic curation may vary depending on the broader political and media environment in which recommendation systems operate, particularly where extremist actors are less prominent in digital information spaces than in France (Caiani & Parenti, 2016; Figeac et al., 2020; Figeac et al., 2021).

Secondly, the study is restricted to the "News" section, which is explicitly curated as an informational space. While this allows for a focused analysis of news visibility, it does not capture how recommendation operates in other parts of the platform. Because the study analyses recommendation outputs rather than the full distribution of available content on the platform, it cannot determine to what extent the observed asymmetries reflect, transform, or amplify the underlying supply of political and journalistic content. Future research should design protocols that enable systematic comparisons across platform segments, between editorially framed zones such as the "News" section and less explicitly curated areas of the platform.

Finally, our models estimate recommendation dynamics based on the immediately preceding video and may not fully capture how YouTube's algorithm integrates longer viewing histories. However, this modelling choice is consistent with our audit design, which isolates sequential recommendation effects under controlled exposure conditions. This limitation should be kept in mind when interpreting continuity effects.

### Conclusion

These findings contribute to the growing literature on recommendation systems by showing that YouTube's NRS operates within a structurally asymmetric visibility regime that cannot be reduced to the aggregation of user histories, interaction patterns, or personalised exposure alone (Yu et al., 2024; Hosseinmardi et al., 2024). By combining cold-start auditing with sequential analysis, this study highlights a form of bias that does not primarily consist in steering users away from news, but in reorganising informational hierarchies within the "News" section itself. In this respect, the article complements existing work on personalisation and on recommendation systems' tendency to privilege engagement-maximising content by showing that recommendation outcomes are shaped not only by user behaviour, but also by the structural baseline from which recommendations begin.

Within this asymmetric regime, political content is not marginalised but selectively sustained. Videos produced by prominent public figures affiliated with ideologically marked parties receive disproportionate visibility within the recommendation outputs, while institutional news outlets, especially from the press and radio sectors, are systematically sidelined. Television broadcasters retain a relative advantage, although still in decline, due to the greater compatibility of audiovisual formats with YouTube's engagement-driven recommendation environment. Continuity effects do not offset this pattern. Rather, they reproduce and stabilise it over successive recommendation steps, reinforcing a visibility structure in which politically salient actors gain prominence more easily than institutionally grounded journalism.

More broadly, these results suggest that structural bias in recommendation systems must be understood in relation to the media and political environments in which platforms operate. While many recent studies emphasise hyper-personalisation and user-level differentiation, our findings show that recommendation systems also operate through asymmetric baseline distributions of visibility that subsequent recommendation sequences tend to sustain rather than correct. In the French case, far-left and far-right actors occupy a particularly prominent position within the visibility structure produced by the "News" recommender. This calls for comparative audits capable of disentangling recommendation effects from the broader media and political environments in which they unfold. Such audits would also contribute to the empirical evidence base required by regulatory frameworks such as the European Union's Digital Services Act for assessing systemic risks associated with recommendation systems.

**Data Availability Statement**

The aggregated data and replication materials supporting the findings of this study are available upon reasonable request. The full underlying dataset is not publicly available because it contains platform-derived materials whose redistribution may be subject to legal, ethical, and technical restrictions.

**Disclosure statement**

The authors report there are no competing interests to declare.

**Funding**

No funding was received.